\documentclass[12pt]{article}
\usepackage{latexsym}
\usepackage{amsmath,amsthm,amssymb}

\def\for{\hbox{for }}

\newcounter{rot}

\def\eps{\epsilon}

  \def\OM{\Omega}

\newtheorem{lemma}{Lemma}
\newtheorem{theorem}{Theorem}

\def\E{\mbox{{\bf E}\;}}

\def\Max{\hbox{MAX}}

\def\vol{\mbox{Vol}}

\def\var{\mbox{{\bf Var}}}

\begin{document}
%
%
%
%
%

\title{Spectral Methods for Matrices and Tensors}

\author{Ravindran Kannan\\
Microsoft Research Labs., India.
\thanks{kannan@microsoft.com}
\thanks{\copyright This is the author's personal version of the work. It is posted here by permission of ACM for personal use, not for distribution. The
definitive version is published in the Proceedings of the ACM Symposium on Theory of Computing, 2010.}
}
\maketitle
\begin{abstract}
While Spectral Methods have long been used for Principal Component Analysis, this survey focusses on work over the last 15 years with three salient features: (i) Spectral methods are useful not only for numerical problems, but also discrete optimization problems (Constraint Optimization Problems - CSP's) like the max. cut problem and similar mathematical considerations underlie both areas. (ii) Spectral methods can be extended to tensors. The theory and algorithms for tensors are not as simple/clean as for matrices, but the survey describes methods for low-rank approximation which extend to tensors. These tensor approximations help us solve Max-$r$-CSP's for $r>2$ as well as numerical tensor problems. (iii) Sampling on the fly plays a prominent role in these methods. A primary result is that for any matrix, a random submatrix of rows/columns picked with probabilities proportional to the squared lengths (of rows/columns), yields estimates of the singular values as well as an approximation to the whole matrix.
\end{abstract}


\section{Introduction}

Spectral methods have been widely used in many areas
for Numerical problems under the name Principal Component Analysis.
The algorithmic use of spectral methods for discrete problems is perhaps more recent. The first point of this survey is to suggest that similar mathematical considerations motivate both discrete and numerical applications.

Our second point is the extension of algorithms to tensors.
Linear Algebra is unique in that it has beautiful theory which also translates to efficient as well as optimal algorithms.
Tensors admit no such comparable theory or algorithms; indeed, some impossibility or hardness results are known for tensors.
This is not our focus. Instead, we want to see what is algorithmically possible with tensors, of course, having to be less ambitious than for matrices. [But we seek provable error bounds.] The second purpose of this survey, then, is to present methods for matrices which extend to naturally to tensors. Two methods surveyed here - the Cut Norm introduced in section \ref{cut-matrices} and
Length-Squared sampling procedure introduced in section \ref{matrix-sampling} both have this flavor.

The third main point of the survey is faster randomized algorithms for large matrices (based on sampling) than traditional numerical algorithms; here length-squared sampling provides a starting point. Besides improving running time through sampling, one operates in a model where the massive matrices cannot be stored in Random Access Memory, but must be read and sampled on the fly. Several newer sampling procedures are also covered in section \ref{sec:relative}.  Many open questions
remain both in regard to algorithms and computational applications. We will list them in the text, but, here we mention two generic challenges to highlight possible directions for research.

{\bf Challenge 1:
Find the spectrum of the web graph.}
The web graph (of hypertext links, for example) is a large ``naturally occurring'' graph. It is directed and so the adjacency matrix is not symmetric. The question is to find the singular values approximately. We will see that the randomized algorithms in the survey could be useful for this. But the provable upper bounds on sample size are too large. Can smaller sample sizes suffice ? We want a certificate on the error bound and confidence (probability of correctness) of the spectrum (say at least of the largest several hundred singular values) as computed for the particular matrix. [In a sense, this is seeking to make a Monte-Carlo algorithm into a Las Vegas one.] The motivation for the particular question is: there is much research on statistical properties of the web graph. But most, pertain to ``local properties'' - degree distribution (which is local in that we find the degree of one vertex at a time), local communities (in a neighborhood of one vertex), etc. The top singular value is a measure of ``global'' correlation. Further, the important notion of pagerank \cite{pagerank} as well as Kleinberg's HITS algorithm \cite{kleinberg} of course have links to the spectrum of the graph.

{\bf Challenge II Better provable Algorithms for Tensors}

The second challenge is: while the methods here make a beginning in dealing with tensors much research remains to be done on algorithms for tensors. We seek better methods for maximizing (approximately) cubic and higher order forms (especially when there are unusually good solutions) and finding low-rank approximations to tensors faster.\\
{\bf Some notation} For a matrix $A$, the Frobenius norm $||A||_F$ is the square root of the sum of the squares of its entries. The spectral norm $||A||_2$ is $\max_{x:|x|=1} |Ax|$. The abbreviation u.a.r stands for ``uniformly at random''. $\epsilon$ will be a positive error parameter.
The material in sections \ref{non-uniform} onwards (which may be read more or less independently of the earlier sections) is covered in greater detail in the monograph \cite{KV09}.
Spectral Graph Partitioning (starting with Fiedler's work \cite{fiedler}) and many other important topics which are well covered elsewhere are not dealt with here.

\section{Approximation of matrices in Cut-norm, applications}\label{cut-matrices}

Instead of the traditional low-rank
approximations to matrices, we start with a form more suited to discrete applications.
For an $m\times n$ matrix, a ``rectangle'' will mean one of the $2^{m+n}$
sets of the form (a subset of rows) $\times$ (a subset of columns). A ``cut
matrix'' is an $m\times n$ matrix with all its entries in some rectangle equal
and all entries outside this rectangle being zero; it is special rank 1 matrix. Our approximation to a matrix is by a sum of cut matrices.
These approximations have
the following desirable properties:
\begin{enumerate}
\item \label{cut-exists} It is very easy to show they exist.
\item \label{cut-alg} They can be found (and this in non-trivial) in polynomial, in fact
in constant time (implicitly) with uniform random sample
of $O(1)$ entries from the matrix.
\item {\bf \label{max-cut}} Using this, one can solve the maximum cut problem to additive error
$\epsilon n^2M$ on $n$ node graphs with maximum edge weight $M$. [This also extends to all MAX-2-CSP's.]
\item {\bf \label{tensors}} Both the existence as well as the algorithmic version
(in constant time) can be extended to tensors and using this,
one can approximately solve MAX-r-SAT for fixed $r$ and other problems.
\end{enumerate}
This is all described in this section.
The main caveat for these methods is the error of $\epsilon n^2M$.
There are many applications where this is not good enough.
In the discrete optimization
setting (say as in Max Cut), if $M=1$, the max cut needs to be $\Omega(n^2)$ or equivalently, the graph needs to be dense for this error to be useful.
Of course, many interesting problems are not dense. A similar situation holds for PCA as well. Section \ref{non-uniform} describes a more amenable error bound for both these areas and how to achieve it using non-uniform sampling and thus address application issues.

$A,B$ will stand for $m\times n$ matrices.
If $S$ is a subset of rows and $T$ a subset of columns,
then, we let
$$A(S,T)=\sum_{i\in S,j\in T}A_{ij}.$$
Define the ``cut norm'' of $||A||_\square$ by
$$||A||_\square=\Max_{S,T} |A(S,T)|.$$
\begin{lemma}\label{cut-exists-lemma}
Assume $|A_{ij}|\leq 1$.
There exist $1/\epsilon^2$ cut matrices whose sum $B$ approximates
$A$ in the sense $$||A-B||_\square\leq \epsilon mn.$$
\end{lemma}
Proof: If $||A||_\square\leq\epsilon mn$, we can take $B=0$. Otherwise, there is
some $S,T$ such that $|A(S,T)|\geq \epsilon mn$. Our first cut matrix in $B$ has
$A(S,T)/|S||T|$ in each entry of $S\times T$ and zero elsewhere.
This subtracts from each $A_{ij},i\in S,j\in T$, their average and it is easy
to show that then $\sum_{i\in S,j\in T}A_{ij}^2$
decreases by at least $\epsilon^2mn$ and so does $||A||_F^2$.
Since $||\cdot||_F\geq 0$, the process must
terminate in at most $1/\epsilon^2$ steps proving the Lemma.

An immediate question is whether such a $B$ can be found in polynomial time.
An affirmative answer was given by Frieze and Kannan \cite{FK99} and indeed, they proved that it can be found in ``constant time'' in a sense to made clear later.

\begin{theorem}\label{cut-alg-thm}\cite{FK99}
For $A$ with $|A_{ij}|\leq 1$ and any fixed $\epsilon>0$,
we can find in polynomial
time a matrix $B$ which is the sum of
$4/\epsilon^2$ cut matrices and satisfies $||A-B||_\square\leq \epsilon mn$.
In fact, given the entries of $A$ in just a u.a.r. rectangle
of size $O^*(1/\epsilon^4)\times O^*(1/\epsilon^4)$,
we can find an implicit description of $B$.
\end{theorem}

From the proof of the Lemma (\ref{cut-exists-lemma}), we see that
it suffices to determine if the cut norm of $A$ is at most $\epsilon mn$ and if not,
find an $S,T$ with $|A(S,T)|\geq \epsilon mn$. The exact
problem is NP-hard. However, for the theorem an approximate version
suffices: find the maximum value of $|A(S,T)|$ to within additive error
$\epsilon mn$. [Really $\frac{\epsilon}{2}mn$, but redefine $\epsilon$ to avoid
putting $/2$ etc.] Reduces to two problems: Max $A(S,T)$ and Max $-A(S,T)$. We describe
the ideas behind a polynomial time algorithm for Max$A(S,T)$ which we may call\newline
{\bf The Maximum Rectangle Problem}:

\begin{itemize}
\item {\bf $S$ gives $T$}

If we know the maximizing $S$, the $T$ to go with it is just the columns of
$A$ whose sum in the $S$ rows is positive.
\item {\bf Estimate Column sums in $S$ rows}

Pick a subset $W$ of $s=O(1/\epsilon^2)$ rows u.a.r.
The sum of each column in the $S$ rows can be estimated (to additive error $\epsilon mn$)
by $\frac{m}{s} \times $ sum in the $W\cap S$ rows.
\item
{\bf Exhaustive Enumeration}

Don't know $S$ or $S\cap W$. But, we can try each subset $\tilde W$ of
$W$ (there are only $2^s$ of them) as a candidate $S\cap W$. For each, find the
set
of columns- $T$ - whose sum in the $\tilde W$ rows is positive.
\item
{\bf Choose best candidate}

For each candidate $T$: Let $S'$ be the rows with
positive sum in the $T$ columns. Take max $A(S',T)$ among all candidate $T$.
\end{itemize}

The Exhaustive enumeration step is inspired by an idea of Arora, Karger and Karpinski \cite{AKK95}.
Why does this algorithm work? We supply only a brief intuition. In the estimating column sums step, if we had the correct $S\cap W$, the only
columns on which we could be wrong about the sign of the column sum in the $S$ rows
are ones where the column sum is close to 0. But these do not contribute much
to $A(S,T)$ anyway. So, one of our candidate $T$ in step 4 is correct in the sense
that $A(S,T)$ is high for the true $S$.
The last step finds the best $T$ among the candidates; we didn't need the true
(unknown) $S$; the best $S$ for each $T$ is just the $S'$.

How do we make this all constant time ? The idea is simple : Pick a u.a.r subset
$\hat S$ of $\hat s=O^*(1/\epsilon^4)$ rows and a u.a.r. subset $\hat T$ of $\hat s$
columns at the outset. In the above algorithm, instead of finding $A(S',T)$ for each of
the $2^{(1/\epsilon^2)}$ candidate $T$ 's, estimate this quantity by
$\frac{mn}{\hat s^2}A(S'\cap\hat S,T\cap\hat T)$, for which we only need to know the entries
of $A$ in $\hat S\times \hat T$. One can show using H\"offding-Azuma inequality that
the estimate is within additive error at most $O(\epsilon mn)$ with high probability.
[The failure probability is at most $e^{-\hat s/\epsilon^2}=e^{-1/\epsilon^2}$ for each candidate;
so by the union bound, whp, there is no failure for any candidate.] Hence the best candidate is
found with asserted error whp.

The approximation $B$ has many algorithmic uses. First consider the Maximum cut problem in
an undirected graph. The problem can obviously be written as (with $A$ = the $n\times n$
adjacency matrix of the graph)
$$\Max_{x\in \{ 0,1\}^n}\quad x^T A (1-x),$$
where $x$ is a column vector and $1$ is the vector of all 1's. We obviously have
$$\left| x^TA(1-x) \; -\; x^TB(1-x)\right|\leq ||A-B||_\square.$$
Here, we see why the cut norm is defined the way it is;
it is the most natural norm for ensuring that for 0-1 vectors
$x,y$, $x^TAy\approx x^TBy$. [It is close to a more traditional ``operator norm''
- namely, it is easy to show that $||A||_\square$ is always
within a factor of 4 of $\Max_{|x|_\infty=|y|_\infty=1} x^TAy$.]
So to get the maximum cut within additive error $\epsilon n^2$, it suffices to
solve $$\Max_{x\in \{ 0,1\}^n} x^T B (1-x).$$ Since $B$ has constant (depending
only on $\epsilon$, not on $n$) rank, $x^TB(1-x)$ is determined by a constant number of
variables, namely the components of $x$ along the space spanned rows/columns of $B$. Thus, we
have reduced the $n$ variable problem to one with a constant number of variables. While
there are many technical difficulties in solving the problem with $B$, conceptually, it is
simple to argue that it can be solved in time exponential in the rank of $B$ alone
by enumeration: Put a fine enough grid in the row/column space of $B$. The number of grid
points is exponential only in the dimension of the space. For each grid point,
the value of $x^TB(1-x)$ is determined. It only remains to know which grid
points correspond to 0-1 $x$ 's. This is an integer program; but its relaxation
Linear Program turns out to suffice for the error we seek.

Indeed, the attractive feature of this line is that
these arguments can be extended in a straight forward manner to solve all dense MAX-2-CSP
problems in polynomial time in a unified manner.
Moreover, this also extends to approximating
tensors in cut norm and that helps us solve all dense MAX-r-CSP problems
for any fixed arity $r$. [Recall: In a MAX-r-CSP problem, one is given a list
of $m$ Boolean functions - $f_1,f_2,\ldots f_m$, each a function of only $r$
of the variables. We have to find a truth setting of all variables which satisfies
as many of the $m$ functions as possible. MAX-r-SAT where each function is the
disjunction of $r$ literals is a central example.]
Indeed, one gets:
\begin{theorem}\label{csp-1}\cite{FK99}\cite{AE02}
Any MAX-r-CSP problem on $n$ variables, where $r$ is fixed, can be solved to additive error $\epsilon n^r$
in constant time for fixed $\epsilon>0$. [The running time depends exponentially on $O^*(1/\epsilon^2)$.]
\end{theorem}
For $r=2$, the area of Property Testing \cite{GGR98}
proved the first such results by combinatorial means,
often by exploiting the structure of particular problems. A flavor of the combinatorial
difficulties is seen from the first problem so attacked - max.cut. by DelaVega \cite{F96}: Akin to
the Estimating columns sums and Exhaustive Enumeration steps of our algorithm for maximizing
$A(S,T)$, the property testing algorithms do the following for max.cut: clearly in a max cut $(S,\bar S)$,
every vertex in $S$ has more edges to $\bar S$ than to $S$. If we just picked a u.a.r. subset
$W$, then try out all possible subsets $\tilde W$ of $W$ as candidate $S\cap W$, we could classify
each vertex by whether it has higher degree into $\tilde W$ or $W\setminus \tilde W$ and hope these would
be respectively $\bar S$ and $S$ vertices. But this does not work and indeed, the property
testing based max cut algorithms work hard to fix this. In our setting, $S,T$ are
subsets of different sets and so are ``decoupled'' and this is what makes
these steps work so simply here.

For general $r>2$, Andersson and Engebretsen \cite{AE02} have given a purely
combinatorial appraoch independently to prove theorem (\ref{csp-1}) too. Our approach to
proving the theorem is based on an extension of cut matrices and norm to tensors which we
describe in the next section.

Another application of the cut norm and approximation is to a version of the Sz\'emeredi Regularity Lemma for graphs. For a graph $G(V,E)$ with edge weights, we denote by $A_G$ the (weighted) adjacency matrix. For two graphs,
$G,G'$, on the same set of $n$ vertices, we define a distance between them by
$d_\square (G,G') =\frac{1}{n^2}\max_{S,T\subseteq V} |A_G(S,T)-A_{G'}(S,T)|$.

From Lemma \ref{cut-exists-lemma}, the following Lemma which is often called the Weak Regularity Lemma can be proved:
\begin{lemma}\label{weak-regularity}\cite{FK99}
The vertex set of a graph $G(V,E)$ with all edge weights equal to 1 can be partitioned into $2^{O(1/\epsilon^2)}$ subsets $V_1,V_2,\ldots $ so that the graph
$G'$ in which for each edge $(i,j)$, with say, $i\in V_r,j\in V_s$, has weight = (number of edges between $V_r$ and $V_s$)/$|V_r||V_s|$ has
$d_{\square }(G,G')\leq \epsilon $
\end{lemma}
Note that $G'$ intuitively behaves like a random graph with edge probabilities given by edge weights in that the number of edges between subsets of vertices would be close to the expected numbers.
A constructive version of the Sz\'emeredi Regularity Lemma was shown in \cite{ADL+}. A simpler spectral algorithm is developed in \cite{FK99-2}.

There is an abstract (and improved) version of this lemma due to Tulsiani, Trevisan and Vadhan \cite{trevisan} from which they are not only able to derive this result, but several others, like the Dense Model theorem of Green, Tao and Ziegler \cite{tao}.
A recent result of Bansal and Williams \cite{BW09} makes progress on the classical problem of complexity of Boolean Matrix multiplication;
they use Lemma \ref{weak-regularity}. We will describe in the next section an application of weak regularity to graph limits.

\newpage

\section{Approximation of Tensors in Cut norm and applications}\label{cut-tensor}
Recall that a $r-$tensor is an $r$ dimensional array $A$ with entries
$A_{ijkl...}$. A ``rectangle'' is a set of entries of the form
$S_1\times S_2\times\ldots S_r$, where $S_t$ is a subset of the $t$ th
index. [For $r=2$, $S_1$ is a subset of rows and $S_2$ a subset of
columns.] $A(S_1,S_2,\ldots S_r)$ is the sum of $A$ 's entries
in the rectangle $S_1\times S_2\times\ldots S_r$. We define the cut norm
$||A||_\square$ exactly as for matrices - it is the maximum over all
rectangles of $|A(S_1,S_2,\ldots S_r)|$.  A cut tensor has the same entry in some
rectangle and is zero elsewhere. Lemma 1 carries over with exactly the
same simple proof as for matrices.
\begin{lemma}\label{cut-exists-2}\cite{FK99}
For a $r-$tensor $A$ with $n_1\times n_2\times \ldots n_r$ entries,
each at most 1 in absolute value,
there exist $1/\epsilon^2$ cut tensors whose sum $B$ approximates
$A$ in the sense $$||A-B||_\square\leq \epsilon n_1n_2\ldots n_r.$$
\end{lemma}
Interestingly, the constructive version also carries over with one
extra twist. The idea for solving the maximum rectangle problem is:
\begin{itemize}
\item Want $S_1,S_2,
\ldots S_r$ so that $A(S_1,S_2,\ldots ,S_r)$ is max.
\item If we knew the maximizing $S_1,S_2,\ldots S_{r-1}$, then the
maximizing $S_r$ consists of the $i$ with \\ $A(S_1,S_2,\ldots ,S_{r-1},i)>0$.
\item We can estimate this by taking random subsets \\ $W_1,W_2,\ldots W_{r-1}$,
trying out all subsets \\ $\tilde W_1,\tilde W_2,\ldots \tilde W_{r-1}$ of the
respective $W_t$ as candidate $S_t\cap W_t$.
\item This gives us many candidate $S_r$. How do we find the best one ?
This needs the extra twist: define a $r-1$ tensor $\tilde A$ by
$$\tilde A_{i_1i_2\ldots i_{r-1}}=\sum_{i\in S_r} A_{i_1i_2\ldots i_{r-1}i}.$$
Now recursively solve the maximum rectangle problem for the $r-1$ tensor. Then
choose the $S_r$ with best answer.
\end{itemize}

The above arguments can be used to show: for a MAX-r-CSP formula $F(x_1,x_2,\ldots ,x_n)$
if we pick a u.a.r. subset $Q$ with $|Q|=q=$ poly$(1/\epsilon)$ variables and solve the ``induced''
MAX-r-CSP problem $F^Q$ on the picked variables (the induced problem contains only those clauses
all of whose literals are the picked variables or their negations), then we have whp:
$$\left| \frac{n^r}{q^r}\Max(F^Q) -
\Max(F)\right|\leq\epsilon n^r,$$
where Max$(F)$ denotes the maximum number of functions in $F$ which can be
simultaneously satisfied.
[$\frac{n^r}{q^r}$ is a natural scaling factor. Note that this is interesting only when
the answer to the whole problem is at least $\Omega(n^r)$. This holds for ``dense'' problems
where there are $\Omega(n^r)$ clauses.] The question arises: what is the best poly$(1/\epsilon)$
in this result? Alon, delaVega, Karpinski and Kannan \cite{AFKK02} prove $O^*(1/\epsilon^4)$ suffices.
\begin{theorem}\label{csp-2}\cite{AFKK02}
Suppose $F(x_1,x_2,\ldots ,x_n)$ is a MAX-r-CSP formula.
 If $Q$ is u.a.r. subset of $q=O^*(1/\epsilon^4)$ of the $n$ variables, then,
for $F^Q$, the induced formula on $Q$, we have with probability at least 99/100,
$$\left| \Max (F)-\frac{n^r}{q^r}\Max (F^Q)\right| \leq \epsilon n^r.$$
\end{theorem}
There are two parts to the theorem: first asserts that \\ $\frac{n^r}{q^r}\Max(F^Q)
\geq\Max(F)-\epsilon n^r$. This is simple: if one just takes the truth assignment
to $\{ x_1,x_2,\ldots ,x_n\}$ which attains $\Max(F)$, then usual facts about
sampling can be used to prove that the SAME assignment to the sampled variables
satisfies (whp) at least $\frac{q^r}{n^r}\Max(F)-\epsilon q^r$ of the clauses of $F^Q$.
The other part is the non-trivial one - the reason is we have to rule out ANY
assignment to the sampled variables from satisfying too many clauses. Indeed, this
raises a basic question:\\
When can we say for a maximization problem that a sampled induced sub-problem gives
a good estimate of the answer to the whole problem ?\\
The non-trivial part is: show that for small problem (the induced one on the sample),
no solution gives an unduly high value. [Traditional sampling arguments tackle the other
part easily.] A situation with a simple answer is bounded Linear Programming: it is easy to see that
if a system of linear inequalities $Ax\leq b; 0\leq x_i\leq 1$ in $n$ variables has a solution, then, for a random
subset $Q$ of the $n$ variables, the induced problem (slightly relaxed) has a solution too:
$A^Qx^Q\leq \frac{q}{n}b+\delta ; 0\leq x_i\leq 1\; \for i\in Q$ where $A^Q$ consists of the columns of $A$ corresponding to the
$Q$ variables. But the converse is also true here using LP duality: if $Ax\leq b$ has no solution with $0\leq x_i\leq 1$, then duality tells us that there is one combination of the inequalities which has no solution; this is equivalent to the existence of a
$u\geq 0$ such that $\sum_j (u^T A)_j^->u^Tb$. Now for this $u$, we can show by traditional sampling that we have: $\sum_{j\in Q} (u^TA)_j^- > \frac{q}{n} (u^Tb)-\delta$  demonstrating that there is no solution to a slight tightening of the sampled LP.
This simple result for LP is used as part of the proof of Theorem \ref{csp-2}.

Another result of a similar flavor about induced subproblems also goes into the proof of Theorem \ref{csp-2} and is worth mentioning independently here. Suppose $A$ is a large $n\times n$ (note: it is square) matrix. If we pick a random subset $Q$ of $[n]$ and look at the induced submatrix $A^Q$ of $A$ on $Q\times Q$, how does the cut norm of $A^Q$ relate to the cut norm of $A$? It is easy to see that $||A^Q||_\square \geq \frac{q^2}{n^2}||A||_\square-\delta$, where $\delta$ is small, since, we could take the subsets $S,T$ of $[n]$ which maximize $|A(S,T)|$ and argue by traditional Statistics that $|A(S\cap Q,T\cap Q)|\geq \frac{q^2}{n^2}|A(S,T)|-\delta$. The theorem below by Rudelson and Vershynin asserts a converse which is harder to prove. It is an improvement of a theorem in \cite{AFKK02} and is proved using some Functional Analysis techniques.
\begin{theorem}\cite{RuVe2007}
Let $\epsilon>0$ and
suppose $A$ is an $n\times n$ matrix with $||A||_F\in O(n)\; ; ||A||_\square\in O(\epsilon n^2)\; ; |A_{ij}|\leq O(1/\epsilon)$. Then if
$Q$ is a u.a.r. subset of $\{ 1,2,\ldots ,n\}$ with $|Q|=q\in\Omega(1/\epsilon^2)$ and $A^Q$ is the $q\times q$ submatrix of $A$ with entries from $Q\times Q$, then
$$E||A^Q||_\square = O(\epsilon q^2).$$
\end{theorem}
{\bf Open Question} \cite{AFKK02} actually proved such a result for $r-$ tensors for any fixed $r$; but their proof required $q\in\Omega^*(1/\epsilon^4)$. Does a result as above with $O(1/\epsilon^2)$ hold for $r-$ tensors ? The issue is that the techniques from Functional Analysis are no more available for $r>2$. (cf. also the next open question on approximating the cut norm has this flavor.)

It is not difficult to show that the problem of finding the cut-norm is MAX-SNP hard by
a reduction from Max-Cut. Interestingly, using a deep result from Mathematics
called Grothendik inequality, Alon and Naor \cite{alon-naor} were able to show:
\begin{theorem}\cite{alon-naor}
The cut
norm of matrices can be approximated to within a factor of 1.782 in polynomial time.
\end{theorem}
Their approach is: the cut
norm problem can be reduced to the following problem:
$$\Max \sum_{i,j} A_{ij}x_iy_j\text{   subject to    } x_i,y_j\in \{ -1,+1\}.$$
This Integer Program has a standard Semi-Definite Programming relaxation, where the $\pm 1$
variables $x_i,y_j$ are replaced by vector variables $u_i,v_j$, required to be of length 1:
$$\Max \sum_{i,j} A_{ij}(u_i\cdot v_j)\text{    subject to   } |u_i|=|v_j|=1.$$
The theorem of Grothendik proves that the optimal value of the SDP is at most a factor
of 1.782 times the optimal value of the integer program. This automatically yields a constant
factor approximation to the value of the integer program. But, finding a rounding procedure
to achieve this was both non-trivial and first developed in \cite{alon-naor}.\\
{\bf Open Problem} Develop $O(1)$ factor for cut norms of $r$-tensors. Note that the natural Semi Definite Program for $r=2$ does not extend to $r=3$.

In the last section, we saw the weak-regularity Lemma and a notion of distance between two graphs.
Borgs, Chayes, Lov\'asz, S\'os, Szegedy and Vesztergombi \cite{BCL+} defined other interesting notions of graph distances and graph limits. They generalize the notion of graph distances to graphs with different numbers of vertices. While the definition in Lemma \ref{weak-regularity} viewed the vertex sets of the two graphs having a fixed 1-1 mapping (labeled vertices), this is no more possible and relabeling of each vertex set as well as mapping one to the other have to be allowed.
We do not give the precise definitions here. Suppose we do the partition of the vertex set as in Lemma \ref{weak-regularity}. Then we could represent each $V_r$ by a compound vertex and put an edge between compound vertices $r,s$ of weight equal to (number of edges between $V_r$ and $V_s$)/$|V_r||V_s|$. Then the Lemma is really saying that the compressed graph and the original are close in some metric. [This intuition needs to be formalized into a definition of distance between graphs.] They then use
Theorem \ref{csp-2} to show:
\begin{theorem}\cite{BCL+}
Let $G$ be a simple graph and $\epsilon,\delta>0$. Then the induced sub-graph of $G$ on a random subset of $2^{1/\delta\epsilon^2}$ nodes is $\epsilon $ close to $G$ with probability at least $1-\delta$.
\end{theorem}
They use this theorem in their extensive work on Graph Limits. Their notions facilitate the understanding of very large graphs which can be viewed as limits; but also can be approximated in the above sense by smaller graphs (with something akin to ``compound vertices''.)
This raises the following somewhat loosely phrased:\\
{\bf Open Question} Can we define a notion of limits for matrices and more generally $r-$ tensors and apply these to derive theorems similar to \cite{BCL+} ?

\section{Non-Uniform Sampling}\label{non-uniform}

Clearly, uniform sampling of rows/columns will not solve all problems. Indeed, if we
have a matrix with just one non-zero row and all other rows were just 0's and we draw
uniform sample of rows, we are likely to see only zeros and miss the all-important row.
Less trivial examples are when only a small number of rows contain significantly higher
absolute value entries than others. From the last sections, we can show that u.a.r. samples
yield an approximation $B$ to the given matrix $A$ with error in cut norm
of at most $\epsilon mnM$, where $M$ is the max absolute value of an entry of $A$.
It can also be shown that with poly$(1/\epsilon)$ u.a.r. samples, we can
make
$||A-B||_2\leq\epsilon \sqrt {mn} M,$
the point being (briefly) $||A-B||_\square = x^T(A-B)y,$ where $|x|=\sqrt m$ and $|y|=\sqrt n$, so
the $||A-B||_2$ error is $1/\sqrt{mn}$ times the $||A-B||_\square$ error.
But this amount of error is not suitable for many applications.

A more useful error
bound is given in Lemma (\ref{EXIST}), for not only matrices, but also tensors.
Completely analogous to the matrix case, we make the following definitions:
For an $r-$tensor $A$, and $r$ vectors $w,x,y,z,\ldots $, $A(w,x,y,z,\ldots )$ is defined
as $\sum_{i,j,k,l,\ldots } A_{i,j,k,l,\ldots }w_ix_jy_kz_l\ldots$. [It is analogous to the
quadratic form $x^TAy=\sum_{i,j}A_{ij}x_iy_j$ for matrices.] The Frobenius norm
of $A$, denoted $||A||_F$ is again the square root of the sum of squares of the
entries. The ``spectral norm'' of $A$ denoted $||A||_2$ is the maximum over all
unit length vectors $w,x,y,z,\ldots$ of $A(w,x,y,z,\ldots )$. A rank-1 $r-$ tensor
is the outer product of $r$ vectors, denoted $w\otimes x\otimes y\otimes z\ldots $
whose $i,j,k,l,\ldots $th entry is $w_ix_jy_kz_l\ldots $. We say that a tensor
has rank at most $k$ if it can be expressed as the sum of $k$ rank-1 tensors.

\begin{lemma}\label{EXIST} \cite{FKKV05}
For any $A$, $\epsilon>0$, there exist
a tensor $B$ of rank at most $1/\epsilon^2$
such that
\begin{equation}\label{A-B-l2}
||A-B||_2\leq \epsilon ||A||_F.
\end{equation}
\end{lemma}
The simple proof of the Lemma will be given shortly.
A polynomial time sampling based
algorithm is also available for producing such an approximation, but the
algorithm given by delaVega, Karpinski, Kannan and Vempala\cite{FKKV05}
is non-trivial.
\begin{theorem}\label{FASTSVD}\cite{FKKV05}
For any $A,\epsilon >0$, we can find a tensor
$B$ of rank at most $4/\epsilon^2$
in time $(n/\epsilon)^{O(1/\epsilon^4)}$
such that with probability at least $3/4$
we have
$$||A-B||_2\leq \epsilon ||A||_F.$$
\end{theorem}

For matrices, traditional singular value decomposition gives us a polynomial
time algorithm, but, we will see a sampling-based algorithm which in essence
can be made constant time after 2 passes through the matrix. In the case of tensors,
no previous polynomial time algorithm was known at all.
Before giving the proofs/algorithms, we will motivate the error bound of $||A-B||_2\leq \epsilon||A||_F$
of (\ref{A-B-l2}) by three application areas.

The first motivating area is Principal Component Analysis (PCA). Here, one often
assumes that the top ``few'' singular values dominate. (In fact, that is in the
first place one of the two justifications for making a low rank approximation. The
other possible motivation for making a low-rank approximation is ``de-noising'' -
where one assumes that the top few singular value components are the real data and
the others are possibly noise- for example in Latent Semantic Indexing \cite{DFLD88}.)

PCA Assumption: The data consists of an $m\times n$ matrix $A$.
The top $k$ singular values $\sigma_1,\sigma_2,\ldots \sigma_k$
contain $1-\epsilon$ of the ``spectrum'', where $k<<m,n$. More precisely,
\begin{equation*}[\text{\bf Strong-PCA}]
\sigma_1^2+\sigma_2^2+\ldots +\sigma_k^2\geq (1-\epsilon)||A||_F^2.
\end{equation*}
We need only a weaker version of this:
\begin{equation*}[\text{\bf Weak-PCA}]
\sigma_1^2+\sigma_2^2+\ldots +\sigma_k^2\geq \Omega(||A||_F^2).
\end{equation*}
Under this assumption, (\ref{A-B-l2}) translates to
a {\it ``relative error} $\epsilon$''.

A second area is Discrete Optimization. As we saw, the max-cut problem can
be solved to additive error $\epsilon n^2M$ for $n-$ node graphs where the edge weights
are all at most $M$. This however is relative error $\epsilon$ only in case
the total of all edge weights is $\Omega(n^2M)$; if $M\leq 1$, this
requires the graph to be dense. This raises the question:

Can we solve non-dense max cut problem to {\it relative error} $\epsilon$ ?
In general, these problems are NP-hard. But an important special case, it turns
out can be solved in polynomial time - namely when the edge weights satisfy the
triangle inequality, as was shown using other methods\cite{FK98}.
A unified polynomial time algorithm using Theorem \ref{FASTSVD}
for this problem and other weighted versions of
MAX-2-CSP problems is developed in \cite{FKKV05}. [In fact, they do this with a weaker condition than triangle
inequality for all MAX-2-CSP problems.]

A third area is tensors. Many algorithms are known and used in practice
for finding low-rank approximations to tensors \cite{kruskal}.
But as remarked earlier, neither
the theory nor the algorithms are anywhere as nice as for matrices. There are solid
reasons - NP-hardness \cite{hastad}, \cite{lek09} and non-uniqueness/existence. But beyond all this, is a basic
question - what is it that we can find provably in polynomial time ? Theorem (\ref{FASTSVD})
seems to be a first step. The algorithm for Theorem (\ref{FASTSVD}) (which we will outline soon) is quite different from
other known heuristics and draws on new uses of sampling in a vein somewhat similar
to the maximum rectangle problem. Also, it turns out that the error bound in
(\ref{A-B-l2}) suffices to tackle MAX-r-CSP problems where the weights satisfy
a natural generalization of the triangle inequality to higher dimensions. Unweighted
dense MAX-r-CSP's are a special case of this.

Proof of Lemma (\ref{EXIST}):
If $||A||_2\leq\epsilon ||A||_F$, then
we are done. If not, there are $w,x,y,z,\ldots $,
all of length 1 such that \\ $A(w,x,y,z,\ldots )
\geq \epsilon ||A||_F$.
Now consider the $r-$dimensional array
\[
B=A-(A(w,x,y,z,\ldots ) w\otimes x\otimes y\otimes
z\ldots
\ldots .
\]
[This is of course basically a rank-1 update.]
It is easy to see that $||B||_F^2=||A||_F^2-
(A(w,x,y,z,\ldots )^2)$. We may repeat on $B$ and clearly this
process will only go on for at most $1/\epsilon^2$ steps.

From the proof of the lemma, it is clear that again, the basic algorithmic question is
to find a $w,x,y,z,\ldots $ all of unit length, maximizing $A(w,x,y,z,\ldots )$ to within
additive error $\epsilon ||A||_F$. We will present the algorithm for tensors later. First, we will tackle matrices by sampling.

\section{Sampling in large matrices}\label{matrix-sampling}

Numerical Analysis gives us sophisticated polynomial time algorithms for many matrix problems to do with spectral analysis. Here, the focus is on using sampling to solve very large matrix problems approximately. First, we look at matrix multiplication. The product of two matrices $A,B$ can be written as
$$AB=\sum_i A_iB^i,$$
where $A_i$ ($B^i$ respectively) is the $i$ column of $A$ (row of $B$, respectively). An immediate thought is to estimate the sum from a random sample of $i$ 's. Consider a random sample of $s$ $i$ 's picked in i.i.d. trials. Let $p_1,p_2,\ldots p_n$ be the probabilities of picking $1,2,\ldots n$ respectively in each trial. If $i_1,i_2,\ldots ,i_s$ are the samples, then $$X=\frac{1}{s}\sum_{t=1}^s \frac{1}{p_{i_t}} A_{i_t}B^{i_t}$$
is easily seen to be an unbiased estimator of $AB$. [I.e., $EX=AB$ entry-wise.] We would like to measure the variance, but this quantity depends on which entry we are talking about. Here, we make a simple-minded, important decision - lets look at the sum of variances of all the entries of $X$. This quantity, which we denote $\var X$ is seen to satisfy:
$$\var X\leq \frac{1}{s}\sum_{i=1}^n  \frac{1}{p_{i}} |A_{i}|^2 |B^{i}|^2.$$
A case of much interest is when $B=A^T$, when this simplifies to
$$\frac{1}{s}\sum_{i=1}^n  \frac{1}{p_{i}} |A_{i}|^4.$$
By Calculus, one can see that this is minimized when the $p_i$ are proportional to $|A_i|^2$. [Indeed it is not hard to show that these $p_i$ are the minimizer of the actual variance, not just the upper bound here.]
This leads to the following probability distribution for sampling the columns of a matrix which turns out to have many nice properties:

{\bf Length Squared Sampling : Pick a column with probability proportional to sum of squares of its entries.}

Length squared sampling was first introduced by Frieze, Kannan and Vempala \cite{FKV-focs98}. Its applications to clustering were studied by Drineas, Frieze, Kannan, Vempala and Vinay \cite{DFKVV04}. The application to matrix multiplication is from \cite{DK2001}. See also \cite{DKM1}.

\cite{FKV-focs98} proves that if we draw a sample of columns according to the length squared distribution and do an SVD on the sampled columns, this gives an low-rank approximation to $A$ with provable error bounds. We state this below (without proof).

\begin{center}
\fbox{\parbox{2.5in}{
\begin{minipage}{2.3in}
\begin{tt}
{\bf Algorithm: Fast-SVD}

\begin{enumerate}
\item[1.] Sample $s$ columns of $A$ from the squared length distribution to form a matrix $C$.
\item[2.] Find $u^{(1)}, \ldots, u^{(k)}$, the top $k$ left singular vectors of $C$.
\item[3.] Output $\sum_{t=1}^k u^{(t)}u^{(t)^T}A$ as a rank-k approximation to $A$.
\end{enumerate}
\end{tt}
\end{minipage}
}}
\end{center}

The matrix $\sum_{t=1}^k u^{(t)}u^{(t)^T}A$ is really just the ``projection'' of $A$ on the space spanned by the $u^{(t)}$ and so the theorem below says that $A$ projected to the top singular space of $C$ (instead of the usual singular space of $A$) is a good low-rank approximation to $A$. [$A_k$ is the best rank $k$ approximation given by SVD.]
\begin{theorem}\cite{FKV-focs98},\cite{FKV04}
The rank-$k$ matrix found by Algorithm Fast-SVD (call it $\tilde{A}$)
satisfies:
$$\E\left(\|A-\tilde{A}\|_F^2\right)\leq \|A-A_k\|_F^2+2\sqrt{\frac{k}{s}}\|A\|_F^2$$
$$\E\left(\|A-\tilde{A}\|_2^2\right)\leq \|A-A_k\|_2+\frac{2}{\sqrt{s}}\|A\|_F^2.$$
\end{theorem}
In fact the kind of error bound in the theorem is optimal in terms of the number of rows sampled; this was shown in \cite{B-Y}.

\cite{FKV-focs98} and \cite{FKV04} in fact apply the sampling once more - to pick a sample of rows of $C$ according to the length-squared distribution. Then, it turns out that fining the SVD of the constant-sized matrix (with the sampled rows of $C$) suffices to give us a low-rank approximation to $A$. But the proof of this is more complicated. The reason is that from the sampled rows of $C$, one gets the right singular vectors of $C$, but only approximately. The error turns out to be bounded by $\epsilon ||C||_F$. [It would be better if the error bound was relative, in terms of the singular values themselves. But length-squared sampling does not give this.] Then for the ``low'' singular values of $C$
(less than $\epsilon ||C||_F$), the approximation is no good. So, one has to throw out these low ones (these are in a sense ``near-singularities'') See \cite{DKM2} for a detailed explanation of the method and some improvements.

An improvement of the error bound, still using length-squared sampling was achieved using sophisticated techniques from the field of Probability in Banach spaces by Rudelson and Vershynin. Their result stated below picks a sample of $s$ rows from $A$, where $s$ is almost linear in a quantity $r$, they call the numerical rank of $A$; $r=||A||_F^2/||A||_2^2$. [Recall the PCA assumptions; under even the weak PCA assumption, $r$ is $O(1)$.] Their error bound is also better in that it involves $||A||_2$, rather than $||A||_F$.
\begin{theorem}\cite{RuVe2007}
Suppose $A$ is an $m\times n$ matrix with numerical rank $r=||A||_F^2/||A||_2^2$. Let $\epsilon ,\delta\in (0,1)$ and $s\in\Omega^* (r/\epsilon^4\delta)$. Let $B$ be a set of $s$ rows of $A$ picked in $s$ i.i.d. trials, each according to length-squared and let $u^{(1)}, \ldots, u^{(k)}$ be the top $k$ right singular vectors of $B$. Then,
$\tilde A=A\sum_{t=1}^k u^{(t)}u^{(t)^T}$ satisfies the following with probability at least $1-2e^{-c/\delta}$:
$$||A-\tilde A||_2\leq \sigma_{k+1}(A)+\epsilon ||A||_2.$$
\end{theorem}
Note also that this is a high probability (with exponential tails) rather than just in expectation.

In another application of length-squared sampling, \cite{vershynin} shows that if we run an iterative equation solver for an overdetermined system of equations with a Kaczmarz iteration (where one uses
a violated equation to modify the current solution) with the added twist that the violated equation is picked according to the length squared distribution, then one gets a guaranteed rate of convergence; the reader is referred to the paper for details.

Length-squared sampling can also be used for tensors and is the basic ingredient in the proof of theorem \ref{FASTSVD}; recall that we needed an algorithm to find for a tensor $A$, the maximum value of $A(w,x,y,z,\ldots )$ to within $\epsilon ||A||_F$. The idea behind the algorithm for this is to imitate the steps of the algorithm for the
maximum rectangle problem:

\begin{enumerate}
\item If we knew the optimizing $x,y,z,\ldots $, then the optimizing $w$ is
easy to find: it is just the vector $A(\cdot,x,y,z,\ldots )$ (whose $i$ th component
is $A(e_i,x,y,z,\ldots )$) scaled to length 1.

\item Now, $A(e_i,x,y,z,\ldots ) = \sum_{j,k,l,\ldots } A_{i,j,k,l,\ldots }x_jy_kz_l\ldots $.
The sum can be estimated by having just a few terms. But, an important question is: how do we make sure the variance is not too high, since the entries can have disparate values ?
\item Length squared sampling works ! [Stated here without proof.]

\end{enumerate}

Achlioptas and McSherry \cite{AM2007} developed a different randomized algorithm for low-rank approximations of matrices - they sample individual entries independently and show using Random Matrix theory that with those on hand, we can get a good approximation to the matrix in spectral norm. Their results also have a bearing on ``Compressed Sensing'' in that they are able to infer something about the whole matrix from a random sample of entries.

\section{CUR: An interpolative low-rank approximation}

We found in the last section an implicit low-rank approximation to $A$; implicit because, the actual approximation needed us to multiply $A$ by the vectors $u^{(t)}$.
In this section, we wish to describe an algorithm to get an explicit approximation of any matrix $A$ given just a sample of rows and a sample of columns of $A$.
Clearly if the sample is picked according to the uniform distribution, this attempt would fail in general. We will see that again the length squared distribution comes to our rescue; indeed, we will show that if the samples are picked according to the length squared or approximate length squared distributions, we can get an approximation for $A$. Again, this will hold for an arbitrary matrix $A$.

First suppose $A$ is a $m\times n$ matrix and $R$ ($R$ for rows) is a $s\times n$ matrix constructed by picking $s$ rows of $A$ in i.i.d. samples, each according to approximate length-squared distribution. Similarly, let $C$
(for columns) be a $m\times s$ matrix consisting of columns picked according to the length squared distribution on the columns. The motivating question for this section is: Can we get an approximation to $A$ given just $C,R$ ? An affirmative answer is given in the theorem below first proved by Drineas and Kannan. Here, one does not need the sampling probabilities to be exactly proportional to length squared; it suffices to have the probability of drawing column $i$ to be at least
its length squared / (c.$||A||_F^2$), where $c$  is a constant. We call this approximate length squared sampling.

\begin{theorem}\cite{DK2003}, \cite{DKM3}
Suppose $C$ (respectively $R$) consists of a sample of $s\geq\Omega^*(k/\epsilon^4)$ columns (respectively rows) of $A$ drawn in
$s$ i.i.d. trials, each according to (approximate) length squared probabilities. Then, from $C$ and $R$, we can find a $s\times s$ matrix $U$ so that
\begin{align*} E||A-CUR||_F&\leq ||A-A_k||_F+\epsilon ||A||_F\\
E||A-CUR||_2&\leq ||A-A_k||_2+\epsilon ||A||_F.
\end{align*}
\end{theorem}

{\bf Open Problem} Improve the dependence of $1/\epsilon^4$ in the theorem.

The approximation of $A$ by the product $CUR$ is reminiscent of the usual PCA approximation based on taking the leading $k$ terms of the SVD decomposition. There, instead of $C,R$, we would have orthonormal matrices consisting of the leading singular vectors and instead of $U$, the diagonal matrix of singular values. The PCA decomposition of course gives the best rank-$k$ approximation, whereas what the Theorem shows for $CUR$ is only that its error is bounded in terms of the best error we can achieve. There are two main advantages of $CUR$ over PCA:
\begin{enumerate}
 \item $CUR$ can be computed faster from $A$ and also we only need to make two passes over $A$ which can be assumed to be stored on external memory.

 \item $CUR$ preserves the sparsity of $A$ - namely $C,R$ are columns and rows of $A$ itself. ($U$ is a small matrix since typically $s$ is much smaller than $m,n$). So any further matrix vector products $Ax$ can be approximately computed as $C(U(Rx))$ quickly.
\item It is an interpolative approximation : unlike SVD, where the singular vectors are linear combinations of $A$'s columns/rows, here $C,R$ are actual columns/rows of $A$. An application illustrates the point. In doing, say, PCA on a ``patient-gene'' matrix in a Biological application (with entry $i,j$ giving the gene expression of gene $j$ for patient $i$), one gets a result which says that ``these few linear combinations of genes/patients are important in explaining the data'', where the linear combinations may involve negative weights as well as positive ones. Instead in CUR, we get a collection of individual genes/patients explaining the data, arguably providing better intuition. See for example \cite{gene-petros} for this kind of application.
\end{enumerate}

The CUR approximation has been extended to tensors as well by Mahoney, Maggioni and Drinesa \cite{tensor-petros}. There are many improvements and applications of CUR - see \cite{CUR-mahoney}. An important modification is given by Sun Xi, Zhang and Falustos \cite{falustos}. They make the observation that the length squared sampling used in the original CUR algorithm may result in a lot of duplicates. They remove duplicates and reweight the unique columns/rows remaining by the square root of the number of copies. They show that with this new reweighting, one gets good approximations. More importantly, they have done empirical studies with Datamining applications to show the effectiveness in terms of space and time with the new approximation.

We also briefly describe an application of CUR to Recommendation Systems by
Drineas, Kerneidis and Raghavan \cite{DKR02}. The central object in a Recommendation System is the customer-preference matrix whose $(i,j)$ th entry is the preference of customer $i$ for product $j$. The basic question is: given a small sample of entries from the matrix (collected customer-product data) how does one make good recommendations to other customers. The idea of using CUR in \cite{DKR02} is a departure from what we have discussed so far. So far, we had (somewhere) the whole matrix and used sampling to infer its properties, it being too large to deal with in full. Here, even collecting the matrix is expensive; we wish to infer its missing entries from the sample we know. But note that the techniques of this survey do apply to this ``reverse engineering'' problem. Indeed, \cite{DKR02} prove that under some assumptions, we can make good recommendations with just a handful of customers' complete preferences (some complete rows) and all customer preferences for a handful of products (columns). Much care has to be taken, since the sampling cannot be assumed to be according to Length-squared.

The above raises a more general question. Can we bring costs of accurate data collection into our measures of efficiency ? We loosely formulate a candidate problem on these lines:

{\bf Open Question} Suppose we wish to solve a large Linear Program : $\max c\cdot x$ subject to $Ax\leq b$. But suppose we have to collect each piece of data - each $A_{ij}$ costs $c/\epsilon^2$ to get with accuracy $\pm\epsilon$. Find an efficient method for data collection + solution to within error $\pm\delta$, where the cost is the weighted sum of the data collection cost plus the running time.

\section{Relative Error Approximation}\label{sec:relative}

If $A_k$ is the best rank $k$ approximation (from SVD), then a natural way to measure error of any rank $k$ approximation is relative to the ``residue'' of the spectrum, namely relative to $||A-A_k||_F$. So, we say a rank $k$ approximation $B$ makes relative error $\epsilon$ if
$$||A-B||_F\leq (1+\epsilon )||A-A_k||_F.$$
From simple examples,
it is easy to see that length-squared sampling does not do this in general.

For multiplicative $(1+\epsilon)$-approximation,
Har-Peled \cite{H-P} gave a linear time algorithm that requires $O(\log n)$ passes over the input matrix. Deshpande and Vempala \cite{DV2006} improved this to $O(k)$ passes using volume sampling. Both these algorithms use adaptive sampling of \cite{DRVW06} as a subroutine. Drineas, Mahoney and Muthukrishnan \cite{DMM} gave a different algorithm, where the sampling probabilities are computed using SVD, that achieves the same approximation ratio but takes more time because of the initial computation of SVD. Finally, Sarl\"os \cite{Sarlos06} gave a linear time 2-pass algorithm for multiplicative $(1+\epsilon)$-approximation that uses a small number of linear combinations of all the rows instead of a subsample; his algorithm which we call isotropic random projection is described below.

First, volume sampling is a generalization of length-squared sampling. We pick
subsets of $k$ rows instead picking rows one by one. The
probability that we pick a subset $S$ is proportional to the
volume of the $k$-simplex $\Delta(S)$ spanned by these $k$ rows
along with the origin. The raw method will give us a factor
$(k+1)$ approximation (in expectation). Incidentally, it also proves that any matrix has $k$ rows whose span contains a such an approximation. Moreover, this bound is tight, i.e., there exist matrices for which no $k$ rows can give a better approximation.

\begin{lemma}\label{KROWS}\cite{DV2006}
Let $S$ be a random subset of $k$ rows of a given matrix $A$
chosen with probability
\[
P_S = \frac{\vol(\Delta(S))^2}{\sum_{T:|T|=k} \vol(\Delta(T))^2}.
\]
Let $\tilde{A}$ be the projection of $A$ to the span of $S$ and let $\tilde A_k$ be the best rank $k$ approximation to $\tilde A$. Then,
\[
\E(||A-\tilde{A}_k||_F^2) \leq (k+1) ||A-A_k||_F^2.
\]
\end{lemma}
More work is needed to convert this to a relative error $\epsilon$ approximation (which we do not describe here.)

Isotropic random projection also gives relative error
approximations to the optimal rank-$k$ matrix with roughly the same time complexity.
Moreover, it makes only {\em two} passes over the input data.

The idea behind the algorithm can be understood by going back to the matrix
multiplication algorithm described in section \ref{matrix-sampling}. There to multiply two
matrices $A, B$, we picked random columns of $A$ and rows of $B$ and thus derived an
estimate for $AB$ from these samples. The error bound derived was additive and this is
unavoidable. Suppose that we first project the rows of $A$ randomly to a low-dimensional
subspace, i.e., compute $AR$ where $R$ is random and $n \times k$, and similarly project
the columns of $B$, then we can use the estimate $ARR^TB$. For low-rank approximation, the
idea extends naturally: first project the rows of $A$ using a random matrix $R$, then
project $A$ to the span of the columns of $AR$ (which is low dimensional), and finally
find the best rank $k$ approximation of this projection. The algorithm is:

\begin{enumerate}
\item Let $l = Ck/\eps$ and $R$ be a random $n \times l$ matrix; compute $B=AR$.

\item Project $A$  to the span of the columns of $B$ to get $\tilde{A}$.

\item Output $\tilde{A}_k$, the best rank-$k$ approximation to $\tilde{A}$.

\end{enumerate}

\begin{theorem}\label{thm:isoRP}\cite{Sarlos06}
Let $A$ be an $m \times n$ real matrix with $M$ nonzeros. Let $0 < \eps < 1$ and $R$ be an $n \times l$
random matrix with i.i.d. Bernoulli entries with mean zero and $l \ge Ck/\eps$ where $C$ is a universal constant.
Then with probability at least $3/4$,
\[
\|A - \tilde A_k\|_F \le (1 + \eps) \|A - A_k\|_F
\]
and $\tilde A_k$  can be computed in two passes over the data
in $O(Ml+(m+n)l^2)$ time using $O((m + n)r^2)$ space.
\end{theorem}

\section{Applications to Clustering, Mixtures}

Spectral methods are widely used for clustering and partitioning problems. Not many worst-case results have been proved; there are exceptions for special classes of graphs like planer graphs \cite{ST2007}.
In general, spectral methods have been proven to work correctly with high probability under some assumptions on the generative model of the data.
One class of such problems is the Planted Problems, where we are given a random graph modified by a ``planted part'' and the objective is to find the planted part. We describe an instance of this.

Consider a graph $G$ which is the union of a purely random graph $G_{n,1/2}$ and an unknown
clique on vertex set $P$, where $p=|P|$ is given.
The problem is to recover $P$.
If $p\geq c(n\log n)^{1/2}$ then with high probability,
it is easy to recover $P$ as the $p$ vertices of largest degree.
Alon, Krivelevich and Sudakov \cite{AKS98}, using
spectral analysis, were able to improve this to $p=\OM(n^{1/2})$.

Let $A_G$ denote the adjacency matrix of $G$.
The spectral approach of \cite{AKS98}
essentially maximizes $x^TAx$ over vectors $x$ with
$|x|=1$, expecting that
the optimal solution is close to $u$,
defined by $u_i=p^{-1/2}1_{i\in P}$, ($u$ is the scaled
characteristic vector of $P$) so that we may
recover $P$ from the optimal solution.

Frieze and Kannan \cite{FK08} define a natural 3-dimensional array $A$
related to the given graph : $A_{ijk}$ will be $\pm 1$
depending on whether the parity of the number of edges
among the vertices $i,j,k$ is odd or even respectively.
They show that as long as $p\in O (n^{1/3} (\log n)^4)$,
the maximum of the cubic form $\sum_{i,j,k} A_{ijk}x_ix_jx_k$
as the vector $x$ varies over the
unit ball is attained close to $u$, so that if we can find this
maximum, then we can recover the clique. However, unlike the
case of the quadratic form, where the maximization was an eignevalue
computation which is well-known to be doable in polynomial time, there are
in general no known polynomial time algorithms for maximizing
cubic forms. So, the existential result does not automatically
lead to an algorithm and this is left as an open question.

{\bf Open Question} Suppose a $n\times n\times n$ array $A$ is constructed
as above from $G_{{n,1/2}}$ plus a planted clique of size $p\in \Omega
(n^{1/3}(\log n)^c)$. Then can we maximize the function $\sum_{i,j,k}A_{ijk}x_ix_jx_k$, $|x|\leq 1,$ even within $O(1)$ factors in polynomial time ?

Brubaker and Vempala \cite{BV09} have generalized this to $r$ tensors, where they show that maximizing over an $r$ tensor whose entries are the parity of the number of edges in $r$ cliques can find hidden cliques of size $n^{1/r}$. The computational question of approximately maximizing the $r-$ary forms is open.

Another class of models is in a sense also planted - there is a hidden partition which dictates probabilities.
For example, McSherry \cite{McSherry01} (following earlier papers) considers a model in which $n$ objects are divided into $k$ clusters ($k<<n$)$T_1,T_2,\ldots T_k$. There is a number $p_{rs}\in [0,1]$ which is the probability of each edge between a vertex in $T_r$ and one in $T_s$.
Edges are chosen independently
and we are given the resulting random graph on $n$ vertices. Our job is to find the partition and $p_{rs}$ of the generating model. This can be summarized as: we are given a 0-1 matrix $A$ and are to find $EA$, where the expectation is entry-wise. \cite{McSherry01} shows that under some technical conditions, spectral methods will yield the answer. Here is a quick idea of the method and its use of the deep results from the theory of random matrices.
The matrix $A-\E A$ has random
independent entries each with mean 0. The following celebrated theorem was first stated qualitatively by the
physicist Wigner and proved by F\"uredi and Komlos\cite{FK81}. See also \cite{Vu05}.

\begin{theorem}\label{wigner}
Suppose $A$ is a symmetric random matrix with independent
(above-diagonal) entries each with standard deviation at most
$\nu$ and bounded in absolute value by 1. Then, with high
probability, the largest eigenvalue of $A-\E A$ is at most $c\nu
\sqrt n$.
\end{theorem}

The strength of this Theorem is seen from the fact that each row of $A-\E A$
is of length $O(\nu\sqrt n)$, so the Theorem asserts that the top eigenvalue amounts only
to the length of a constant number of rows; i.e., there is almost no correlation among
the rows (since the top eigenvalue $=\max_{|x|=1}\|(A-\E A)x\|$ and hence the higher
the correlation of the rows in some direction $x$, the higher its value).
Thus one gets whp an upper bound on the spectral norm of $A-EA$:
$$\|A-\E A\|\leq c\nu\sqrt n.$$
Now, we can (approximately) find $EA$ by doing SVD on $A$ with the help of the following simple lemma.

\begin{lemma}\label{AMtoF}\cite{AM2007}
Suppose $A,B$ are $m\times n$ matrices with rank$(B)=k$. If $\hat A$ is the best rank $k$
approximation to $A$, then
$$\|\hat A-B\|_F^2\leq 5k\|A-B\|^2.$$
\end{lemma}
While these ideas are clean, it turns out that they only help cluster ``most'' points correctly. The others are corrected in a messy ``clean-up'' phase. There has been progress on clustering in generative models: \cite{AFKM2001},\cite{Dasgupta2005},\cite{DHKM07}. But the messiness of the clean-up phase haunts the field and raises the following:

{\bf Open Problem} Clean up the clean-up phase of clustering algorithms for generative models (or dispense with it).

Another well-studied clustering problem has to do with learning mixtures of Gaussians and other probability densities. We only describe the part of this area which has to do with spectral algorithms. A provable connection to spectral method was struck by Vempala and Wang \cite{Vempala2004}. They proved an elegant result that given samples from a mixture of $k$ spherical Gaussians, the $k-$ dimensional SVD subspace of the matrix whose rows are the samples contains all the centers. With the space of centers in hand, one can project to that subspace and learn in it. Extensions of this were given in \cite{Kannan2008} and \cite{Achlioptas2005}.
Two interesting variants of PCA have been proposed recently by Brubaker and Vempala - isotropic PCA \cite{Brubaker2008} and robust PCA \cite{Brubaker2009a} which tackle Gaussian mixture learning problems not amenable to standard PCA.

While we do not go into the subject of Spectral Partitioning of graphs, we briefly mention that there are many ways to partition nodes given edge weights which are to be treated as pairwise similarities between vertices. A well-used method is to normalize first each row sum to be 1, then find the second largest eignevalue and corresponding eigenvector of the the stochastic matrix. Then we partition the vertex set into 2 subsets: those with coordinate in the second eigenvector and those with low coordinates. The cut-off can be chosen. This algorithm and its variations are widely used \cite{SM}. Not many proofs of error bounds are known though. \cite{ST2007} prove bounds for planer graphs. \cite{KVV04} prove that if one repeats this partitioning procedure on the subgraphs, then we can ensure that the graph is ultimately split into parts of high conductance with not too much edge weight ``wasted'' between different parts. \cite{KM} prove better bounds for planer graphs.

{\bf Acknowledgements} I am grateful to Alan Frieze and Santosh Vempala for their collaboration on work reported here and to Santosh also for his comments on the manuscript. Thanks to all my coauthors.

\newcommand{\etalchar}[1]{$^{#1}$}
\providecommand{\bysame}{\leavevmode\hbox to3em{\hrulefill}\thinspace}
\providecommand{\MR}{\relax\ifhmode\unskip\space\fi MR }
\providecommand{\MRhref}[2]{%
  \href{http://www.ams.org/mathscinet-getitem?mr=#1}{#2}
}
\providecommand{\href}[2]{#2}


\end{document}